\documentstyle[epsfig,12pt]{article}
\newcommand {\epem}        {\mathrm{e}^+\mathrm{e}^-}
\newcommand {\epemto}      {\mathrm{e}^+\mathrm{e}^-\rightarrow}
\renewcommand {\aleph}     {\sc {ALEPH}\rm}
\newcommand {\delphi}      {\sc {DELPHI}\rm}
\newcommand {\lll}         {\sc {L3}\rm}
\newcommand {\opal}        {\sc {OPAL}\rm}
\newcommand {\lep}         {\sc {LEP}\rm}
\newcommand {\lplm}        {$\ell^{+}\ell^{-}$}
\newcommand {\lsp}         {$\widetilde{\chi}^{0}_{1}$}
\newcommand {\chapm}       {$\widetilde{\chi}^{\pm}_{1}$}
\newcommand {\chapmp}      {$\widetilde{\chi}^{\pm}_{2}$}
\newcommand {\chap}        {$\widetilde{\chi}^{+}_{1}$}

\newcommand{\pma}[2]{\mbox{\raisebox{-0.6ex}
{$\stackrel{\scriptstyle \;+\; #1}{\scriptstyle \;-\; #2}$}}}
\begin{document}
\begin{titlepage}
\begin{flushright}
WISC-EX-96-345\\
June 11, 1996
\end{flushright}
~~\\
\vspace{0.1cm}
\begin{center}
{\Large SEARCHES AT \lep\   1.5 } \\  
\vspace{1cm}
%
% Your NAME
%
\renewcommand{\thefootnote}{\fnsymbol{footnote}}
Jane NACHTMAN\footnote{Supported by the
US Department of Energy, grant number DE-FG0295ER40896.}\\
%
% Your AFFILIATION and ADDRESS
%
Department of Physics, University of Wisconsin, Madison, WI~53706, USA\\
\end{center}
\vspace{7cm}
\begin{abstract}

The energy upgrade at \lep\  
\rm allows new regimes to be explored in 
the search
for physics beyond the Standard Model.  The searches for new
physics using the \aleph , \delphi , \lll , and \opal\ data are described,
and the results are presented. 

\end{abstract}          

\end{titlepage}

\section{Introduction}

As \lep\ increases energy from the Z peak to the W pair threshold, new 
opportunities in searches for
physics beyond the Standard Model are presented.
%New physics could hold the 
%explanation for the current puzzles of the origin of dark matter, or the 
%anomalous measurement of $\mathrm R_{b}$.  
The \lep\ 1.5 run, which consists
of nearly $\mathrm 6~pb^{-1}$ of data per experiment at energies of 
130 and 136~GeV, gives
an opportunity to explore unprecedented energy regimes in  $\epem  $ 
collisions for evidence of compositeness, fourth generation fermions, and 
Supersymmetry.

\section{Experimental Aspects of Searches at \lep }

Topologically, many signatures for new physics at \lep\ have similar 
components and similar detector requirements.  Missing energy is at the 
heart of many searches for new physics, and so accurate energy measurement and
detector hermeticity are essential.  Tracking and calorimetry for jet 
reconstruction, and identification of leptons and photons are necessary to 
reconstruct signatures of new physics.

As \lep\ moves away from the Z peak, the physics opportunities for search 
analyses increase, and the search conditions change.  Signatures for new 
physics have lower background from Z decays, and the more exotic 
backgrounds from four fermion processes, WW and $\mathrm ZZ^{*}/\gamma$, 
 play 
an increased role.  Although the cross sections are lower than those 
at \lep\ 2, the event topologies are very similar to those of 
signal processes.  The highest production cross section comes
from two photon events.  However, these events normally have characteristics 
very different from those of signatures for new physics, so they are 
mostly easily eliminated from search analyses.

Four fermion processes are an important background for searches at LEP,
especially for searches for Supersymmetry and the Higgs boson.  An
understanding of the rate and characteristics of these events is thus 
essential.  At \lep\ 1.5, the dominant four fermion process comes from the
production of an on-shell Z boson and a virtual photon.  
%The event topologies 
%are characterized by one pair of high-mass ($\mathrm M_{f\bar{f}} 
%\approx M_{Z}$) and one pair of low-mass  ($\mathrm M_{f\bar{f}}^{2} \approx
%Q^{2}$ of photon propagator) fermions.
The \opal\ collaboration have reported an excess of four fermion final state 
events with high multiplicity~\cite{opal-4f}.  
The \aleph\ Collaboration~\cite{aleph-4f}
in an analysis 
sensitive to both high and low multiplicity four fermion final states, observe
rates consistent with Standard Model expectations obtained with the 
{\sc FERMISV} \rm program~\cite{fermisv}.
The results are given in 
Table ~\ref{table-fourfermion}.

\section{Searches for Compositeness}

Evidence for compositeness may appear in direct searches for new particles, 
or through anomalies in Standard Model processes.

Results of searches for single and pair production of excited leptons are
reported by \\
\aleph\ ~\cite{aleph-lstar},
 \delphi\ ~\cite{delphi-lstar},
 and \lll\ ~\cite{l3-lstar}.    
Excited leptons usually decay
radiatively, which gives a signal topology for pair production of two leptons 
and two 
energetic photons.  For excited
neutrinos, the radiative decay may be forbidden, and in that case, the excited
neutrino decays to a lepton and a W, and the signal topologies are determined 
by the W decays (i.e., combinations of leptons and jets with rates
determined by the branching ratios).  Efficiencies for the search are typically between
45 and 60\%, with the $\tau^{*}$ search having a lower efficiency than 
$\mathrm e^{*}$ and $\mathrm \mu^{*}$ searches.  The production cross section 
for pair 
production of excited leptons can be as large as  8~pb, and combined with the high 
efficiency of the analyses, limits were set on excited lepton masses to 
nearly the kinematic limit, as shown in Table~\ref{table-lstar}.  Single 
production of excited leptons results in events with two leptons and a photon, 
$\epem \rightarrow \ell \ell^{*} \rightarrow \ell \ell \gamma$.  The production
cross section for this process is low ($\leq$ 0.2~pb), except for excited 
electrons, which can have a t-channel enhancement of the cross section. The 
results of the
searches can be interpreted as limits on the coupling between excited leptons
and their partners, which is related to the compositeness scale, as a function
of the mass of the excited lepton.

% put four fermion table here to move it back
\begin{table}
\begin{center}
\begin{tabular}{l|l|l||l|l|l}  \hline
    & \multicolumn{2}{|c||}{\opal} & \multicolumn{3}{c}{\aleph}  \\ \cline{2-6}
  &                                      %space  
  $\mu^{+}\mu^{-}\mathrm{q\bar{q}}$  &   
  $\mathrm{e^{+}e^{-}q\bar{q}}$      &
  \lplm $\mathrm{q\bar{q}}$          & 
  $\mathrm{\nu\bar{\nu}f\bar{f}}$    & 
  \lplm \lplm                      \\  \hline \hline
  \sc{FERMISV} \rm                 & 
  0.55\pma{0.04}{0.04}$\pm$0.07    &   % opal mumuqq 
  0.63\pma{0.06}{0.06}$\pm$0.10    &   % opal eeqq
  2.50$\pm$ 0.18                   &   % aleph llqq
  2.38$\pm$ 0.14                   &   % aleph nunuff
  1.62$\pm$ 0.08                   \\ % aleph llll
  bkg.                             & 
  0.07\pma{0.06}{0.03}$\pm$0.01    &                   % opal mumuqq 
  0.09\pma{0.07}{0.04}$\pm$0.03    &                   % opal eeqq
  0.14\pma{0.11}{0.02}             &                   % aleph llqq
  0.04\pma{0.08}{0.02}             &                   % aleph nunuff
  0.01\pma{0.08}{0.01}             \\ \hline \hline   % aleph llll
Total Exp.     &
\multicolumn{2}{c||}
{1.34\pma{0.10}{0.08}$\pm$0.17} &  % opal total
\multicolumn{3}{c}
{6.69\pma{0.29}{0.24}}  \\         % aleph total
Observed       &  
\multicolumn{2}{c||}
{6}       &  % opal total observed  
\multicolumn{3}{c}
{5}      \\  % aleph  
\hline
\end{tabular}

\caption[.]
{\it Four fermion events observed by the \opal\ \it and \aleph\ \it 
collaborations, and 
Standard Model expectations given by \sc{FERMISV}\it  and background Monte 
Carlo.
\opal\  \it numbers include statistical and systematic uncertainties, 
and \aleph\ \it 
statistical and systematic uncertainties are combined in quadrature.
\label{table-fourfermion}}

\end{center}
%\end{table}

% excited lepton table  
%\begin{table}
\begin{center}
\begin{tabular}{l|c|c|c|c|c}  \hline 
  & $\mathrm e^{*}$ & $\mathrm \mu^{*}$ & $\mathrm \tau^{*}$ 
& $\mathrm \nu^{*} (\ell W)$ & $\mathrm e^{*} (\nu \gamma)$
 \\
\hline \hline
%          e*         mu*       tau*       nu*(lw)      nu* (nugamma)
\aleph  & 65.1    &  65.4   &   64.7   &    -      &     -            \\
\hline
\delphi & 63.5    &  63.5   &   63.1   &    -      &     -            \\
\hline
\lll    & 64.7    &  64.9   &   64.2   &   57.3    &    61.2          \\
\hline
\end{tabular}
\caption[.]
{\it Mass limits in GeV for excited leptons from the search for 
$\epem \rightarrow \ell^{*} \ell^{*} \rightarrow \ell \gamma \ell \gamma$.
Results are given assuming radiative decays, except for neutrinos, where 
the decay $\mathrm \nu^{*}\rightarrow \ell W$ has also been considered.
 \rm
\label{table-lstar}}

\end{center}
\end{table}

Events with two coplanar photons can be examined for evidence of compositeness
through contact interactions or existence of excited electrons produced in 
the t-channel.  New physics is detectable through deviations 
in differential cross sections from the Standard 
Model predictions.  Limits can be set on
QED cutoff parameters from these comparisons.   Also, if the 
$\mathrm ee^{*}\gamma$ and $\mathrm ee\gamma$ couplings are assumed to be 
equal, a limit on the mass of the excited electron can be derived.  These 
results are given in Table~\ref{table-multiphoton} for the 
\aleph\ ~\cite{aleph-gg}, 
 \lll\ ~\cite{l3-gg}, 
and  \opal\ ~\cite{opal-gg} 
analyses.

% coplanar photon (QED cutoff terms)  and m(e*) limits
\begin{table}
\begin{center}
\begin{tabular}{l|c|c|c}  \hline 
 & \multicolumn{2}{c|}{QED cutoff parameters} & 95\% CL \\  \cline{2-3}
 &            $\Lambda_{+}$ (GeV) & $\Lambda_{-}$ (GeV) & $\mathrm M(e^{*})$  \\
\hline \hline
\aleph\   &        169            &     132             &        137        \\
\hline 
\lll\     &        131            &     167             &        129        \\
\hline
\opal\    &        152            &     143             &        147        \\
\hline
\end{tabular}
\caption[.]
{\it Limits derived from the study of coplanar photon events: QED cutoff
limits, $\Lambda_{+}$ and $\Lambda_{-}$, and 95\% CL limit on the mass of the 
excited electron, in GeV.
\rm
\label{table-multiphoton}}

\end{center}
\end{table}

\section{Fourth Generation Leptons}
        Searches for fourth generation leptons are reported by 
\lll\ ~\cite{l3-L}.
Charged heavy leptons are pair produced and then decay via a W to their 
neutral partner, which is stable, massive,  and invisible.  The experimental 
topologies are determined by the W branching ratios, as each event will 
have two W's and two massive invisible particles.  The topologies are a 
pair of leptons, a lepton and jets, or jets, all with missing energy and 
mass.  No events are selected in the \lll\  analysis, while  0.9 
events are expected from Standard Model background.  
From the predicted production 
cross section of 
1-4 pb, charged heavy leptons are excluded up to 62 GeV, if the mass difference
between the charged and neutral lepton is high enough.

        Pair production of neutral heavy leptons which decay via a W to 
a lower generation lepton is also considered.
Topologies determined by the W branching ratios have at least 
two leptons and jets.  In the \lll\  analysis, 0.9 events were expected in 
the electron channel and 0.3 in the muon channel, and none were selected 
in the data.  Heavy Dirac neutrinos are excluded for masses 
up to 58.9 GeV, and Majorana neutrinos up to 48.1 GeV.

        \aleph\  reports an analysis for t-channel production of a 
single heavy neutrino and first generation neutrino~\cite{aleph-lstar}.  
Isosinglet 
heavy neutrinos are predicted in models such as the Seesaw Model.   The heavy 
neutrino decays to a W and first generation lepton, and so the topologies 
include an electron and the products of the W decay.  The degree of mixing 
between the fourth and first generation determines the cross section, 
so the result of the analysis is a limit on the amplitude of the mixing, 
as a function of the mass of the heavy neutrino.  Although \lep\ 1 limits on 
the mixing amplitude are not exceeded, the mass range is increased.

\section{Single Photon Searches }

Pair production of neutral stable particles, such as neutrinos or 
the lightest supersymmetric particle, 
can not be detected except in the case where 
an initial state photon is detected.  Measurement of the single photon 
rate was used to set limits on $\epem \rightarrow $ \sc{XX}, \rm where \sc{X}
 \rm is any 
neutral stable particle.  The dominant Standard Model process,
$ \epem \rightarrow \nu\nu\gamma$, is an irreducible background for this
signature.
The cross sections reported 
by \aleph\ ~\cite{aleph-gg}~, 
 \delphi\ ~\cite{delphi-g}~, 
and \opal\ ~\cite{opal-gg}  
are consistent with expectations for 
$ \epem \rightarrow \nu\nu\gamma$.

%The cross sections reported 
%by \aleph\ , \delphi\ , and \opal\  are listed in Table ~\ref{table-singlephoton}.

%\begin{table}
%\begin{center}
%\begin{tabular}{l|c|r|r}  \hline 
%  &  $\sqrt{s}$ (GeV)  &  $\sigma (\epem \rightarrow XX\gamma) (pb)$ &  $\sigma (\epem \rightarrow \nu\nu\gamma) (pb)$ \\
%\hline \hline

%\aleph & 130 & 9.6 $\pm$ 2.0 & 10.7 $\pm$ 0.2  \\
%       & 136 & 7.2 $\pm$ 1.8 &  9.1 $\pm$ 0.2   \\
%\hline
%\delphi & 130 + 136 & 13.9 $\pm$ 1.0 & 13.2 $\pm$ 1.0 \\
%\hline
%\opal  & 130 & 3.3 $\pm$ 1.3 & 5.0  \\
%       & 136 & 7.2 $\pm$ 2.0 & 4.2 \\
%\hline
%\end{tabular}
%\caption[.]
%{\it Measured cross section for $\epem \rightarrow XX\gamma$, and 
%Standard Model prediction for $\epem \rightarrow \nu\nu\gamma$.  The 
%prediction depends on the fiducial cuts used in the analysis. \rm
%\label{table-singlephoton}}

%\end{center}
%\end{table}

\section{Supersymmetry}

The Minimal Supersymmetric extension of the Standard Model predicts a 
supersymmetric partner for all Standard Model particles.  This results in a
spectrum of new particles.  The supersymmetric partners of the charged bosons, 
$\mathrm\widetilde{W}^{\pm}$ and $\mathrm\widetilde{H}^{\pm}$,
mix to form two states of charginos, \chapm\ and \chapmp\ .  
 The neutral bosons,
$\mathrm \widetilde{\gamma}, \widetilde{H}^{0}$, and $\mathrm\widetilde{Z}^{0}$ , 
mix to form 
neutralinos, $\widetilde{\chi}^{0}_{1}$, $\widetilde{\chi}^{0}_{2}$, 
$\widetilde{\chi}^{0}_{3}$, and $\widetilde{\chi}^{0}_{4}$ .  The 
supersymmetric partners
of the fermions of the Standard Model are the sleptons, $\mathrm \widetilde{e}, 
\widetilde{\mu}, \widetilde{\tau}$, and $\widetilde{\nu}$, and the squarks, $\mathrm 
\widetilde{q}$. 

R parity, a new quantum number having the value of -1 for supersymmetric 
particles and +1 for Standard Model particles, is introduced
in order to prevent fast proton decay.  In the analyses
reported here, R parity conservation is assumed.  
Experimental consequences of this 
assumption are that SUSY particles are produced in pairs, and the 
lightest SUSY particle, the LSP, does not decay.  The lightest neutralino, 
$\widetilde{\chi}_{1}^{0}$, is massive and neutral, and escapes detection.  The signatures
for Supersymmetry are events with missing energy and mass, due to the 
presence of the LSP.

Experimentally, Standard Model processes can mimic the signal when
the events have  missing 
energy from initial state radiation, neutrinos, or particles at low angles 
which are not detected.  The search is made more difficult when the mass 
difference between the next-to-lightest supersymmetric particle (the NLSP) and 
the LSP is small, since this leads to events which are difficult to detect, 
and kinematically resemble the two-photon background.

The \aleph\ ~\cite{aleph-susy}~, 
 \delphi\ ~\cite{delphi-cha,delphi-chi}~,
 \lll\ ~\cite{l3-susy}~,  
and  \opal\ ~\cite{opal-cha,opal-stop}  
searches use kinematic 
properties of the signal in order to differentiate it from the background.
The signal, with 
missing energy and mass from the LSP, has large transverse momentum 
imbalance.  Typical
Standard Model backgrounds, in particular two-photon processes, have 
missing momentum parallel to the beam
direction, and thus low transverse momentum.  Transverse imbalance can also be 
exploited by the quantity of the scalar sum of the momenta perpendicular to the 
thrust axis in the transverse plane, which is used in slepton searches 
against the $\gamma\gamma\rightarrow\tau^{+}\tau^{-}$ background.
Another distinguishing characteristic is the distribution of energy in the 
detector, as $\gamma\gamma$ events will have a large fraction of the total 
visible energy concentrated at low angles.  The background from
$\epem\rightarrow\mathrm\gamma Z\rightarrow\mathrm q\bar{q}\gamma $ 
can have missing energy due to the ISR photon, a neutrino from a heavy quark
semileptonic decay, or mismeasurement.  A useful property of the signal 
is the isolation of the missing momentum vector, projected into the 
plane transverse to the beam, which allows separation of 
$\mathrm q\bar{q}\gamma$ events from the signal.
Using kinematic quantities such as these, the 
\lep\ analyses are able to obtain very low background expectations, while 
retaining reasonable signal efficiencies, for mass differences (NLSP-LSP) above
5~GeV.  
%5~$\mathrm GeV/c^{2}$.  

\subsection{Results of Searches for Supersymmetric Particles}

Charginos are theoretically favored to be light, and considered to 
have the highest discovery potential 
since the predicted cross sections are high ($\sim$5-20 pb) and the 
experimental signatures are clear.  The cross 
section, however, can be decreased by t-channel interference if the sneutrino 
is light.  If the sneutrino is lighter than the chargino, two body decays 
dominate (\chapm$\rightarrow\ell^{\pm}\widetilde{\nu}$).  Typically, however, the
chargino decays to a neutralino and a W, so the topologies are determined by
the W branching ratios.  The experimental topologies are: a pair of acoplanar 
jets, jets and a lepton, and acoplanar leptons, all with missing energy and 
mass.  Efficiencies reported for chargino detection are as high as 74\%, for 
a mass difference (\chap\ - \lsp\ ) of 30 GeV, with an expected background
of approximately one event.  No evidence for chargino 
production was found, and mass limits were set nearly up to the kinematic 
limit, as shown in Table ~\ref{table-mass}.

% susy mass limits
\begin{table}
\begin{center}
\begin{tabular}{c|c|c|c|c}  \hline 
                                         & \aleph\       &    \delphi\    
&    \lll\    &   \opal\   \\
\hline \hline
$\mathrm M({\widetilde{\chi}^{\pm}_{1}}$) & 67.8          &     66.8       
&    65.0     &   65.4     \\ 

\hline
\end{tabular}
\caption[.]
{\it 95\% CL limits on the chargino mass in GeV, assuming 
$\mathrm\Delta M(\widetilde{\chi}^{\pm}_{1}-\widetilde{\chi}^{0}_{1})\geq $10~GeV, 
and 
$\mathrm m_{\widetilde{\nu}}~\geq ~200~GeV$.
\label{table-mass} }

\end{center}
\end{table}

The search for neutralinos is complementary to the chargino search, 
since the highest neutralino production cross sections ($\sim$ 7~pb)
occur in the higgsino 
region, where the chargino selection efficiency is lower due to the near mass 
degeneracy of the chargino and neutralino.
Neutralino searches at
\lep\ 1.5 focused on associated production of $\widetilde{\chi}^{0}_{2}$\lsp\ 
where 
$\widetilde{\chi}^{0}_{2}\rightarrow$\lsp\ $\mathrm Z^{*}$;  thus the neutralino signatures are 
determined by the Z branching ratios, leading to topologies of acoplanar jets
or leptons, with missing energy and mass.  Efficiencies as high as 30\%
are attainable when the mass difference ($\widetilde{\chi}^{0}_{2}$ - \lsp\ ) is 
10-15~GeV.
No evidence for neutralino 
production was found by the \lep\ experiments.

Squarks are usually expected to be too heavy, and the cross section too 
small, to be of interest for current
searches at \lep\ , except under special circumstances.  In the case of 
the $\tilde{t}$, 
large Yukawa coupling leads to large off-diagonal terms in the mass matrix,
causing mixing between the left and right  states.  Thus, there can 
be a large mass splitting between the two $\tilde{t}$ states, so
 that the lightest $\tilde{t}$ could be 
detectable at \lep\ energies. The
production cross section for $\epemto\tilde{t}\bar{\tilde{t}}$ depends on the mixing 
angle.
The topology expected for stop decays is a pair of acoplanar jets, as each stop
squark decays to a charm quark and a neutralino.  Typical search efficiencies 
are between 50 and 64\%, for a mass difference ($\mathrm \tilde{t}$-\lsp) of 
$\mathrm 30~GeV/c^{2}$.  
Searches by \aleph\ ,
\delphi\ , \lll\ , and \opal\  
found no candidates, and limits were placed on the stop mass as a function of 
neutralino mass and mixing angle.  In the most optimistic case, where 
$\mathrm\Theta_{mix}$=0 and the production cross section is highest, 
limits on the $\tilde{t}$ mass of 52-57 GeV are set by the \lep\ experiments.

The topology for pair production of sleptons, 
$\epemto\widetilde{\ell}^{+}\widetilde{\ell}^{-}\rightarrow\ell^{+}$\lsp$\ell^{-}$\lsp\, is a pair of acoplanar leptons, accompanied by missing energy 
and mass. Typical search efficiencies are about 70\%, for $\mathrm \widetilde{e}$'s
 and $\widetilde{\mu}$'s, when the \lsp\ is $\sim$ 30~GeV lighter than the slepton.  
Efficiencies fall when the mass difference between the slepton and \lsp\ 
decreases, and are generally lower for $\widetilde{\tau}$'s.
No candidates were selected in the \aleph\  
and \lll\  
analyses, which expected approximately one event from background processes,
and limits can be placed on the slepton mass as a function of 
neutralino mass.  Because the production cross sections for smuons and staus 
are low ($\sim $0.5~pb), the limits set at \lep\ 1 are not improved.
However, for selectrons, the production cross section can be 
enhanced by t-channel neutralino exchange, if the neutralino is light
enough and contains a significant gaugino component.  
In that case, improved limits can be set on the mass of the selectron,
to 53~GeV.

%\begin{table}
%\begin{center}
%\begin{tabular}{l|c|c|c|c}  \hline 
%& \multicolumn{3}{c}{ Efficiency($\mathrm\Delta 
%M(\widetilde{\ell},LSP)=5,30~GeV/c^{2}$)}                        & Exp. Bkgd. \\
%  &  $\mathrm\widetilde{e}$  & $\widetilde{\mu}$   &  $\widetilde{\tau}$ & (events)   \\ 
%\hline \hline
%\aleph\ &  40,72\% & 45,75\% & 7,54\% & 1.3\\
%\hline
%\lll\ &    54,69   &  48,67  & 6,56   & 0.9 \\
%\hline
%\end{tabular}
%\caption[.]
%{\it Selection efficiencies and expected number of background events in 
%the \aleph\ \it and \lll\ \it slepton analyses.  \rm
%\label{table-slepton}}
%\end{center}
%\end{table}

\begin{figure}
\begin{center}
\mbox{\epsfig{file=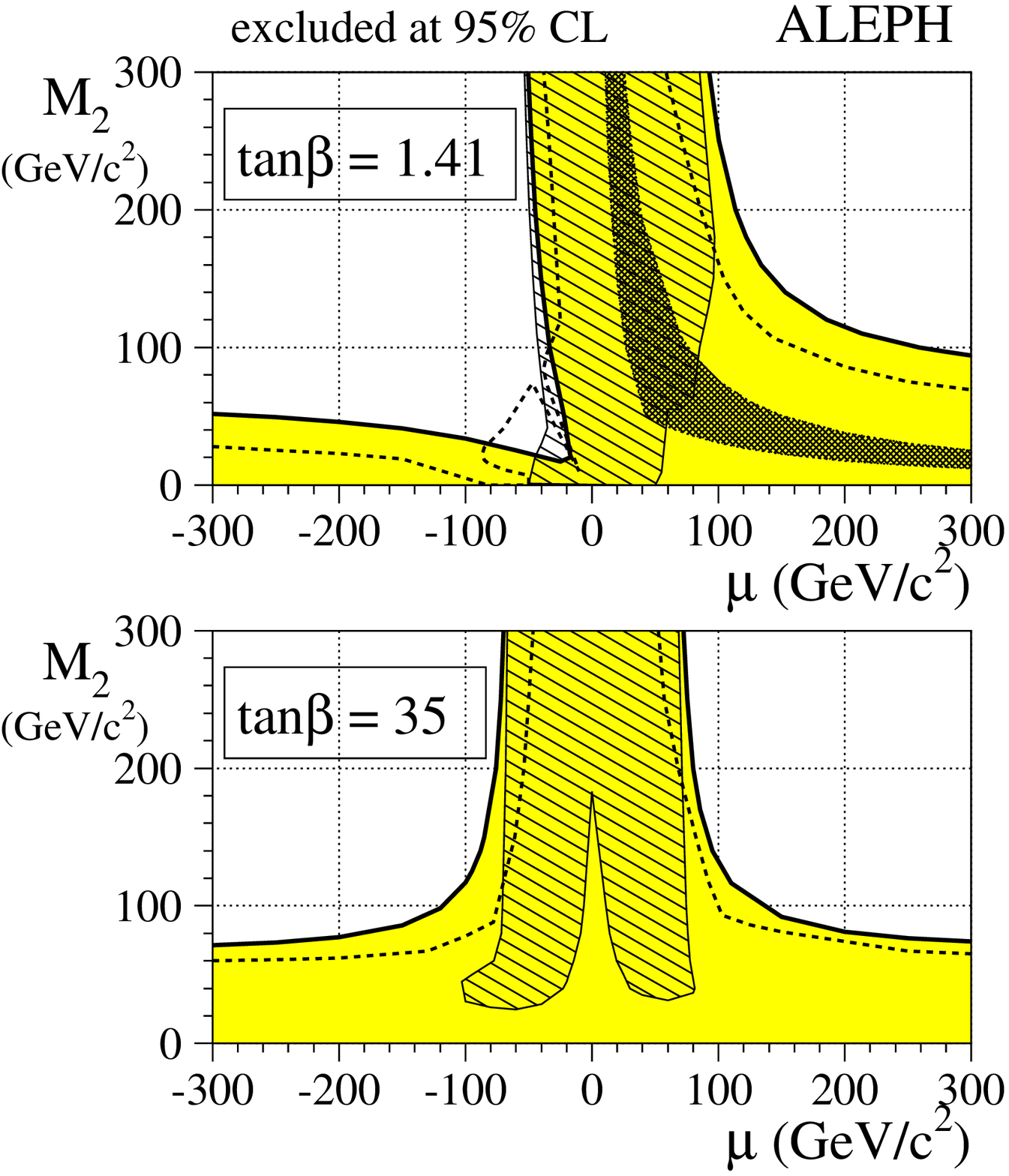,height=11cm%
,bbllx=1mm,bblly=25mm,bburx=170mm,bbury=235mm}}
\end{center}
\caption[.]{\em  Regions in the $\mathrm \mu -M_{2}$ plane excluded by
chargino and neutralino searches, for tan$\beta$ = 1.41 and 35, and 
$m_{\widetilde{\nu}}$=500 GeV. The shaded 
region is excluded by chargino searches and the hatched region by
neutralino searches at LEP 1.5.  The dashed lines indicate the LEP 1 limit.
\label{fig-mu_m2} }
%\end{figure}

% chi mass
%\begin{figure}
\begin{center}
\mbox{\epsfig{file=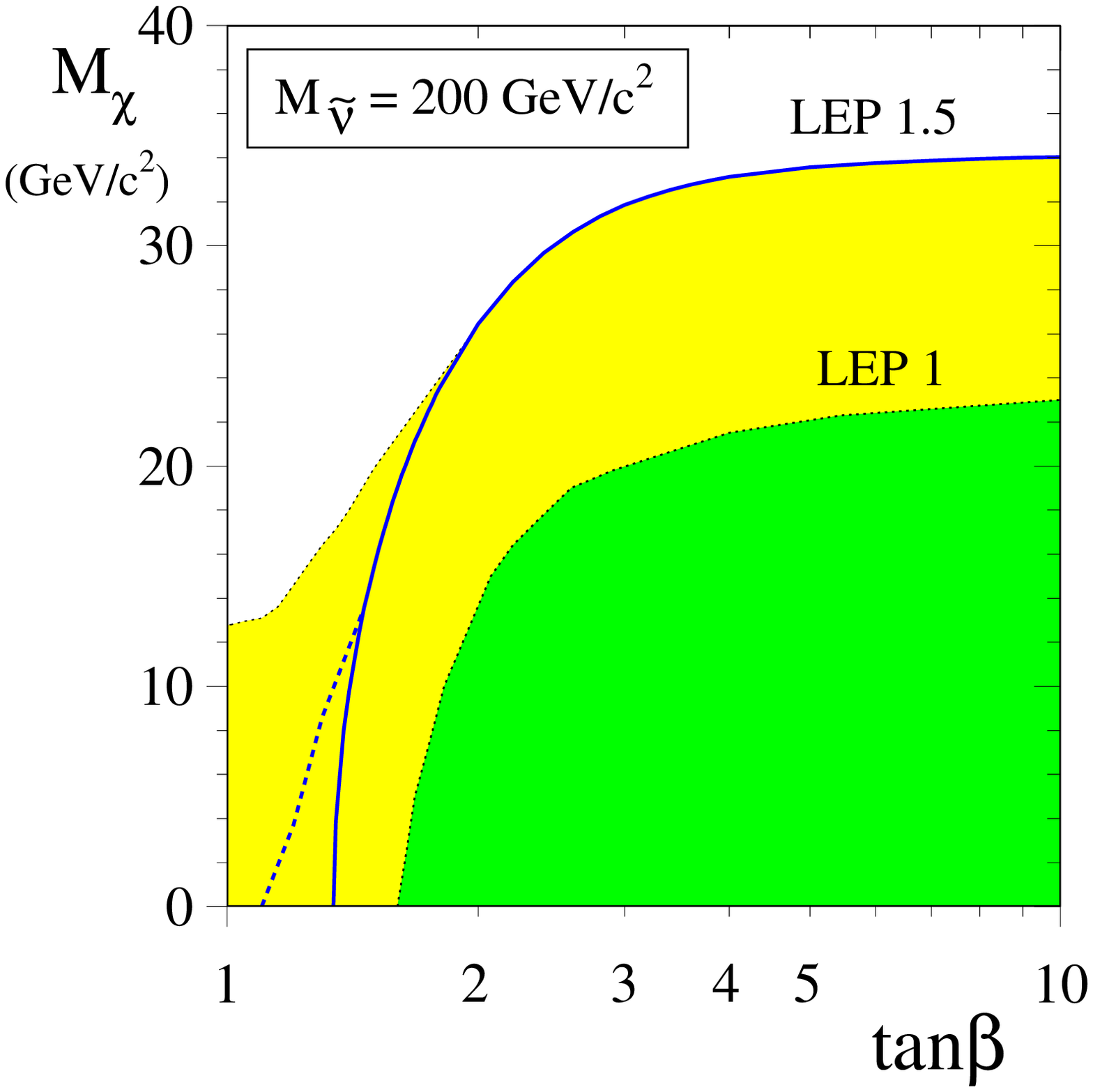,height=7cm%
,bbllx=1mm,bblly=55mm,bburx=165mm,bbury=225mm}}
\end{center}
\caption[.]{\em Lower limit on the mass of the lightest neutralino as a
function of tan$\beta$, for $m_{\widetilde{\nu}}=200$~GeV, derived by ALEPH .
The dark shaded region 
is the 
LEP 1 limit.  The solid line is the limit from chargino searches, and 
the dashed line from neutralino searches at LEP 1.5.  The light shaded region
is the mass of the lightest neutralino excluded by the 
combination of the LEP 1 
and LEP 1.5 limits.
\label{fig-mchi} }

\end{figure}

The limits on the masses of the supersymmetric particles can be interpreted 
in the parameter space of the MSSM.  Chargino and neutralino masses depend 
on three 
parameters; $\mathrm tan\beta$, $\mu$, and $M_{2}$.  Assuming a heavy 
$\widetilde{\nu}$ ($\mathrm m_{\widetilde{\nu}} \geq 200~GeV$), improved limits on $
\mu$ and $\mathrm M_{2}$ for two values of tan$\beta$ are shown in 
Figure~\ref{fig-mu_m2}.

An interesting result is derived by \aleph\  for the combination 
of \lep\ 1 and \lep\ 1.5
 results~\cite{aleph-mchi}.  
The mass of the lightest neutralino is not 
bound for all values of
tan$\beta$ from the \lep\ 1 or \lep\ 1.5 results, if taken separately.  
However, if the results are combined, they complement each other such that 
the mass of the lightest neutralino is restricted to be $\geq$~12.8~GeV, for 
all tan$\beta$, as shown in Figure~\ref{fig-mchi}.

\section{Conclusion}

The recent increase of energy at \lep\ has allowed new regions of parameter 
space to be explored in the search for new physics.  
No discoveries are reported in 
searches for 
compositeness,
fourth generation leptons, and Supersymmetry, but improved exclusion limits 
are set.

%%%%%%%%%%%%%%%%%%%%%%%%%%%%%%%%%%%%%%figures
% mu-M2

\end{document}